\documentclass[10pt]{article}
\usepackage[margin=1.15in]{geometry}
\usepackage[square,numbers]{natbib} 
\usepackage{graphicx,psfrag,color}
\usepackage{bbm}
\usepackage{dsfont}
\newcommand{\1}[1]{\mathds{1}
}
\usepackage[mathscr]{eucal}
\usepackage{amsmath,amsthm,amssymb}
\usepackage[colorlinks=false]{hyperref}
\def\tag{\renewcommand{\theequation}}
\newcommand{\Int}[2]{{\displaystyle \int_{ #1}^{ #2}}}
\newcommand{\lemmref}[1]{{\rm Lemma \ref{lemma:#1}}}
\newcommand{\ode}[2]{{\displaystyle \frac{\mbox{$d #1$}}{\mbox{$d #2$}}}}
\newcommand{\half}{\mbox{$\frac12$}}

\renewcommand{\L}{\boldsymbol{\mathsf L}}
\newcommand{\calc}{\mathscr C}

\newcommand{\N}{\boldsymbol{\mathsf{N}}}
\newcommand{\M}{\boldsymbol{\mathsf{M}}}
\newcommand{\A}{\boldsymbol{\mathsf{A}}}
\newcommand{\B}{\boldsymbol{\mathsf{B}}}
\newcommand{\propref}[1]{{\rm Proposition \ref{proposition:#1}}}
\newcommand{\Bp}{\begin{proposition}\begin{sl}}
\newcommand{\EP}[1]{\end{sl}\label{proposition:#1}\end{proposition}}

\newcommand{\Eqref}[1]{{\rm (\ref{eq:#1})}}
\newcommand{\Bt}{\begin{theorem}\begin{sl}}
\newcommand{\Bl}{\begin{lemma}\begin{sl}}
\newcommand{\Bc}{\begin{corollary}\begin{sl}}
\newcommand{\Ec}{\end{sl}\end{corollary}}
\newcommand{\ET}[1]{\end{sl}\label{theorem:#1}\end{theorem}}
\newcommand{\EL}[1]{\end{sl}\label{lemma:#1}\end{lemma}}
\newcommand{\theoref}[1]{{\rm Theorem \ref{theorem:#1}}}
\newcommand{\Frac}[2]{\displaystyle{\frac{\displaystyle{#1}}{\displaystyle{#2}}}}
\newcommand{\be}{\begin{equation}}
\newcommand{\ba}{\begin{array}}
\newcommand{\ea}{\end{array}}
\newcommand{\ee}{\end{equation}}
\newcommand{\eeq}[1]{\label{eq:#1}\end{equation}}
\newcommand{\real}{{\mathbb R}}

\newcommand{\ed}{\end{document}}

\newcommand{\R}{\ensuremath{\mathbb{R}}}

\newcommand{\e}{e}
\newcommand{\diver}{\mathop{\mathrm{div}}}

\newcommand{\norm}[1]{\lVert #1 \rVert}
\newcommand{\V}[1]{\boldsymbol{#1}}
\newcommand{\T}[1]{\mathbb{#1}}
\renewcommand{\d}{{\rm d}}

\newtheorem{theorem}{Theorem}[section]

\newtheorem{lemma}[theorem]{Lemma}
\newtheorem{proposition}[theorem]{Proposition}
\newtheorem{remark}[theorem]{Remark}
\newtheorem{corollary}[theorem]{Corollary}
\title{Nonlinear Stability Analysis of a Spinning Top \\ with an Interior Liquid-Filled Cavity}
\author{Giovanni P. Galdi\thanks{Department of Mechanical Engineering and Materials Science, University of Pittsburgh, USA}\ \ \ \& \ Giusy Mazzone\thanks{Department of Mathematics and Statistics, Queen's University, Canada}}
\date{}
\begin{document}
\maketitle
\begin{abstract}Consider the  motion of the coupled system, $\mathscr S$, constituted by a (non-necessarily symmetric) top, $\mathscr B$, with an interior cavity, $\mathscr C$, completely filled up with a Navier-Stokes  liquid, $\mathscr L$. A particular steady-state motion $\bar{\sf s}$ (say) of $\mathscr S$, is when $\mathscr L$ is at rest with respect to $\mathscr B$, and $\mathscr S$, as a whole rigid body, spins with a constant angular velocity $\bar{\V\omega}$ around a vertical axis passing through its center of mass $G$ in its highest position ({\em upright spinning top}). We then provide a complete characterization of the nonlinear stability of $\bar{\sf s}$ by showing, roughly speaking, that $\bar{\sf s}$ is stable if and only if $|\bar{\V\omega}|$ is sufficiently large, all other physical parameters being fixed. Moreover we show that, unlike the case when $\mathscr C$ is empty, under the above stability conditions, the top will eventually return to the unperturbed upright configuration.
\end{abstract}
\renewcommand{\theequation}{\arabic{section}.\arabic{equation}}
\section*{\large Introduction}
The stability  of an upright spinning top is a renowned,  classical problem in  rigid body dynamics. More precisely, the top is a rigid body, $\mathscr B$, that moves while keeping one of its points, $O$, fixed at all times. Among the many motions that $\mathscr B$ can execute, particularly interesting is the one where $\mathscr B$ spins with constant angular velocity, $\bar{\V\omega}$,  around the vertical axis, ${\sf a}$, passing through $O$ and its center of mass $G$ at its highest position, and coinciding with one of the principal axes of inertia $\V e_{\mathscr B}$ (say). The 
above-mentioned problem consists  exactly in studying the stability properties of such rotational motion. It is then well known that, if $|\bar{\V\omega}|$ is ``sufficiently large" and the principal moment of inertia around $\V e_{\mathscr B}$ is a maximum, then the motion is stable and the perturbed motion will be  a combined precession-nutation  around ${\sf a}$  \cite[Example 9.7C]{Pars}.    
\par
In this paper we investigate the analogous  question when the top has an interior cavity, $\mathscr C$,  entirely filled with a viscous liquid, $\mathscr L$. In this case the basic motion is the same  as above, with $\mathscr L$ at rest with respect to $\mathscr B$ and the coupled system $\mathscr S:=\mathscr B\cup\mathscr L$ rotating as a whole rigid body around the direction ${\sf a}$ coinciding with $\V e_{\mathscr S}$. We denote by $\bar{\sf s}$ this particular unperturbed steady-state. Mainly due to its important technological applications (\cite{ChB} and the references therein), this problem has attracted the attention of a number of applied mathematicians, especially from the Russian School. As a matter of fact, Sobolev \cite{Sob} was the first to {\color{black} give sufficient conditions for the stability} of $\bar{\sf s}$,  in the case when the liquid is {\em inviscid} (Euler),  within the {\em linearized} approximation, namely, when all nonlinear terms in the relevant equations are entirely disregarded. 
Precisely, let $A,B$ and $C$ be the principal moments of inertia of  $\mathscr S$ with respect to $O$ with $C$ moment of inertia about $\V e_{\mathscr S}$. In \cite{Sob} it is then shown that a sufficient condition for {\em linear} stability is that
\be
C>M:=\max\{A,B\}\,,\ \ \bar{\omega}^2>\frac{\beta^2}{C-M}\,,
\eeq{00}
where $\beta^2$ is the material constant defined in \eqref{eq:notation}.
Successively,  by a  clever but {\em formal}~\footnote{No existence theory is provided for the underlying initial-boundary value problem.} application of  the classical Liapunov method, Rumyantsev \cite{Rum} proved that the requests in \Eqref{00} in fact ensure full {\em nonlinear} stability when $\mathscr L$ is viscous (Navier--Stokes). Instability results were further furnished by  Chernus'ko \cite{Che} and Smirnova \cite{Sm1,Sm2} by means of a formal asymptotic expansion obtained for small relevant Reynolds number, and  suitable symmetry assumptions on the shape of the cavity. More recently, Kostyuchenko {\em et al.} \cite{Kal} have shown, among other things, that, if $\mathscr L$ is a Navier-Stokes liquid, under the assumption \Eqref{00}$_1$, the request in \Eqref{00}$_2$  is, basically, {\em necessary and sufficient}  for the stability of $\bar{\sf s}$. More precisely, \Eqref{00} ensures stability, whereas the conditions  
\be
C>M:=\max\{A,B\}\,,\ \ \bar{\omega}^2<\frac{\beta^2}{C-M}\,,
\eeq{001}
implies that $\bar{\sf s}$ is unstable. However, again as in \cite{Sob}, the result in \cite{Kal} is proved for the {\em linearized} set of equations only; see also \cite{KK,Yu}.
\par
Over the last few years, the present authors and their associates have started a systematic and stringent analysis of the motion of a rigid body with an interior cavity entirely filled with a viscous liquid \cite{DGMZ,Ga,GMpendulum,GMM1,GMM,Ma,Ma2,MPS,MPS1}. In particular, in \cite[Section 8]{GMM}, it has been shown in a rigorous way  that condition \Eqref{00} guarantees the full nonlinear stability of the steady-state $\bar{\sf s}$ in a very large class of perturbations (weak solutions). However, the instability counterpart of this result requires a condition that is more restrictive than that stated in \Eqref{001}. 
\par
The objective of this paper is to furnish a rigorous, complete and detailed analysis of {\em nonlinear asymptotic stability} of the steady-state motion $\bar{\sf s}$ of $\mathscr S$. The method we employ is based on the study of the spectral properties of the relevant linear operator --obtained by linearizing the full nonlinear operator around $\bar{\sf s}$-- in combination with a ``generalized linearization principle" of the type introduced in \cite{GMpendulum}; see \theoref{1}. {\color{black} One of the key ingredient of our proof is hypothesis (H5) on the nonlinear operator of our evolution equation \Eqref{1.7}. Such assumption ensures that the null set of the relevant linear  operator is contained in the manifold of equilibria for \Eqref{1.7}, and it defines a stable center manifold which attracts nearby trajectories. }
We are thus able to show that the conditions stated in \Eqref{00} ensure the nonlinear asymptotic stability of $\bar{\sf s}$ in a suitable regularity class of perturbations (\theoref{6.1}). Here, by ``asymptotic stability" we mean that $\bar{\sf s}$ is stable in the classical sense of Liapunov and, moreover, the generic perturbed motion with initial data in an appropriate  neighborhood of $\bar{\sf s}$ will converge, exponentially fast, to a steady-state where the top continues to spin in the upright position. This ``stabilizing effect" is just due to the presence  the liquid since, as we noted earlier on, if the cavity is empty the perturbed motion of the top is  precession-nutation-like around the vertical axis through the fixed point $O$. In addition, we provide two sufficient conditions for the nonlinear instability of $\bar{\sf s}$. Precisely, we show that $\bar{\sf s}$ is unstable if either \Eqref{001} holds or else the principal moment of inertia $C$ is not a maximum (Theorem \ref{th:instability}). The latter result, in its full generality, is entirely new. 
\par
The plan of the paper is as follows. In Section 1 we furnish a mathematical formulation of the stability problem along with some preparatory results. In Section 2,  \theoref{1}, we prove a``generalized linearization principle" for the trivial solution of an abstract evolution problem in a Banach space. 
Unlike the classical ``linearization principle" \cite[Theorem 5.1.1]{henry} where, for stability, the spectrum of the linearization, $\L$ (say), must have eigenvalues with positive real part,  our theorem allows  0 to be in the spectrum of $\L$, on condition that it is a semi-simple eigenvalue. We also show that, if $\L$ has at least one eigenvalue with negative real part, then the trivial solution is unstable; see \theoref{UST}. This result would follow also from \cite[Theorem 5.1.5]{henry}; however, we give an independent and simple proof for the sake of completeness. {\color{black} In Section 3, we show that the linearization around $\bar{\sf s}$ obeys all regularity and spectral properties of the abstract operator $\L$. 
The study of the location of the spectrum, $\sigma$, of the linearization is then carried in Section 4. 
In Section 5 we prove that the relevant nonlinear operator in our stability problem satisfies the assumptions of the analogous operator in the abstract setting. 
Finally, in Section 6 we are able to apply our abstract results to study the stability properties of $\bar{\sf s}$. 
}
\setcounter{equation}{0}
\section{\large Formulation of the Problem and Preliminary Considerations}

{\color{black} In the following and for the rest of this paper, boldfaced lower-case letters will denote vector fields, whereas boldfaced capital letters will denote operators and tensor fields. }

\begin{figure}[h]
\centering
\psfrag{O}{$O$}
\psfrag{1}{$\V e_1$}
\psfrag{2}{$\V e_2$}
\psfrag{3}{$\V e_3$}
\psfrag{G}{$G$}
\psfrag{B}{$\mathscr V$}
\psfrag{L}{$\mathscr C$}
\psfrag{g}{\color{black} $\V g$}
\includegraphics[width=.4\textwidth]{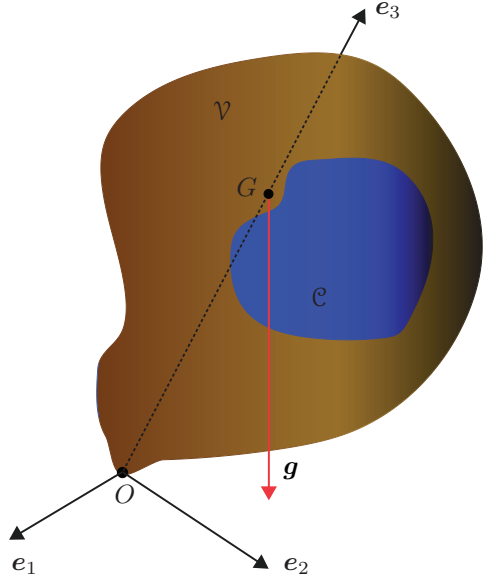}
\caption{A liquid-filled spinning top.}\label{fig:top}
\end{figure}

Let $\mathscr S:=\mathscr B\cup\mathscr L$  the {\color{black} physical system} constituted by a rigid body $\mathscr B$ possessing an interior cavity entirely filled with a Navier-Stokes  liquid $\mathscr L$. {\color{black} We denote by $\mathscr V$ and $\mathscr C$ the spatial region occupied by $\mathscr B$ and $\mathscr L$, respectively. In mathematical terms, $\mathcal V$ and $\mathscr C$ are  two bounded domains of $\R^3$  with $\bar{\mathscr C}\subset \mathscr V$. 
We suppose that  $\mathscr C$ is of class $C^2$.} We  assume that $\mathscr S$ moves under the action of the gravity, $\V g$, while keeping one of its points, $O$, fixed at all times with respect to an inertial frame (see Figure \ref{fig:top}). 
Denote by  $\mathbb I$ the inertia tensor of $\mathscr S$ with respect to $O$,  $\V e_1, \V e_2,$ and $\V e_3$ its (ortho-normalized) eigenvectors and  $A,B,$ and $C$  corresponding eigenvalues ({\em principal moments of inertia}). Moreover, let $\mathcal F:=\{O,\V e_1,\V e_2,\V e_3\}$ denote the principal frame of inertia.  
  We shall suppose that the center of mass, $G$, of $\mathscr S$ belongs to the axis $\vec {O\V\, e_3}$, and  $G\neq O$. Also, we orient $\mathcal F$ in  a way that the only non-zero coordinate, $\ell$, of $G$ in $\mathcal F$ is positive. Thus, the equations governing the motion of $\mathscr S$ in the body-fixed frame $\mathcal F$ are given by \cite{Ma2,MR}
\begin{equation}\label{eq:motion}
\begin{split}
&\left.\begin{split}\smallskip
&\V v_t+\V v\cdot \nabla \V v+\dot{\V \omega}
\times \V x+2\V \omega\times \V v =\nu\Delta \V v-\nabla p 
+g \,\V \gamma
\\
&\diver \V v=0
\end{split}\right\}\text{ in }{\color{black}\mathscr C}\times (0,\infty)\,,
\\
&\left.\begin{array}{ll}\smallskip\mathbb I\cdot(\dot{\V \omega}-\dot{\V a}) +\V \omega\times \mathbb I\cdot ({\V \omega}-{\V a})
=\beta^2 \V e_3\times \V \gamma
\\ 
\V{\dot \gamma}+\V \omega\times \V \gamma=\V 0 
\\ \end{array}\right\}  \text{ in }(0,\infty)\,,\\
&\hspace*{3mm}\V v=\V 0\qquad \text{ on }{\color{black}\partial\mathscr C\times (0,\infty)}\,.
\end{split}
\end{equation}
Here $\V v,\rho,  \mu\equiv \rho\nu$ and $\V \omega$  denote, respectively,  relative velocity, density, 
  shear viscosity coefficient of $\mathscr L$ and angular velocity of $\mathscr B$. Moreover, 
\begin{equation}\label{eq:notation}
\beta^2:=M\ell g, \;\; 
\V a:=-\rho\,\mathbb I^{-1}\cdot \int_{\mathscr C}\V x\times \V v,\;\;
p:={\sf p}-\frac \rho2 (\V \omega \times \V x)^2,
\end{equation}
where ${\sf p}$ is the pressure field of $\mathscr L$, 
 while 
$M$ is the total mass of  $\mathcal{S}$, and $g=|\V g|$. Finally,  
$\V \gamma:=\V g/g$ stands for the direction of the gravity, which, in 
the  frame $\mathcal{F}$, is time-dependent. Notice that
\be
\V \gamma^2(t)=1\,,\ \ \mbox{for all $t\ge 0$}\,.
\eeq{gam}
{\color{black} 
\par
As it is immediately checked,  \eqref{eq:motion} admits a class of time independent (steady-state) solutions  $(\bar{\V v}(x),\bar{\V\omega},\bar{\V\gamma})$, characterized by the conditions 
\be
\bar{\V v}\equiv\V0\,,\ \bar{\V\omega}=\lambda\,\bar{\V \gamma}\,,\ \ \bar{\V\gamma}=-\V e_3\,,\ \ \lambda\in\R\setminus\{0\}\,.
\label{s}
\end{equation}
From the physical viewpoint, such  solutions describe those motions of $\mathscr S$, where the liquid is ``frozen" in the cavity and $\mathscr S$ rotates as a whole rigid body around $\V e_3$ with constant angular velocity $\V\omega$, and its center of mass at its highest position (``{\em upright spinning top}"). 

The main objective of this paper is to characterize the stability properties of these motions.

To this end, let 
\be
(\bar{\V v}+\V v, \bar{\V \omega}+\V\omega, \bar{\V\gamma}+\V z),
\eeq{pert}
be a generic ``perturbed motion" around (\ref{s}). Then, in view of \eqref{eq:motion} the ``perturbation" $(\V v, \V\omega, \V z)$ satisfies the following set of equations
\begin{equation}\label{eq:perturbation}
\begin{array}{ll}\medskip
\left.\begin{array}{ll}\smallskip
\V v_t+\dot{\V\omega}\times \V x+\V v\cdot\nabla \V v+2(\V\omega-\lambda\,\V e_3)\times \V v=\nu\Delta \V v-\nabla p
\\
\diver \V v=0
\end{array}\right\}\ \ \text{on }{\mathscr C}\times (0,\infty)\,,\medskip
\\\left.\ba{ll} \T I\cdot (\dot{\V \omega}-\dot{\V a})+\V \omega\times \T I\cdot (\V \omega-\V a)
-\lambda\,\V e_3\times\T I\cdot(\V\omega-\V a)-\lambda\,C\,\V\omega\times\V e_3  =\beta^2\V e_3\times \V z 
\medskip \\
\V{\dot z}+\V \omega\times\V z-\V \omega\times \V e_3-\lambda{\V e_3}\times \V z=\V 0\ea\right\}\ \mbox{in $(0,\infty)$}\,,\medskip
\\
\hspace*{2mm} \V v=\V 0\quad \text{on }\partial \mathscr C\times(0,\infty)\,.
\end{array}\end{equation}
Because of \Eqref{gam}. the perturbation field $\V z$ must satisfy the constraint
\be
\V z(t)\cdot\V z(t)-2\V z(t)\cdot\V e_3=0\,,\ \ \mbox{all $t\ge 0$}.
\eeq{zet}
 However, dot-multiplying {\color{black}\eqref{eq:perturbation}$_4$} a first time by $\V z$, a second time by $-\V e_3$ and summing side by side the resulting equations, we find
$$
\ode{}t\left(\V z\cdot\V z-2\V z\cdot\V e_3\right)=0\,,\ \ \mbox{for all $t\ge 0$}\,.
$$
As a consequence, \Eqref{zet} is equivalent just to require that the initial data $\V z(0)$ satisfies:
\be 
\V z(0)\cdot\V z(0)-2\V z(0)\cdot\V e_3=0\,.
\eeq{zet0}
\par
Our next objective is to rewrite \eqref{eq:perturbation} as an evolution equation in an appropriate Hilbert space. To this end,  let
\[
L^2_\sigma(\mathscr C):=\{\V v\in L^2(\mathscr C):\; \diver \V v=0\quad \text{in }\mathscr C,\;\V v\cdot\V n=0\quad\text{on }\partial \mathscr C\}
\]
and define the Hilbert space 
\begin{equation}\label{eq:X}
X:=\{\V u=(\V v,\V \omega,\V z)^T\in L^2_\sigma(\mathscr C)\oplus\R^3\oplus\R^3\}, 
\end{equation}
endowed with the inner product 
\[
\langle\V u_1,\V u_2\rangle=\int_{\mathscr C}\rho\, \V v_1\cdot\V v_2\; \d V+\V \omega_1\cdot\T I\cdot\V \omega_2+\V z_1\cdot\V z_2
\] 
and associated norm 
\[
\norm{\V u}:=\sqrt{\langle\V u,\V u\rangle}.
\]
We then introduce the  operators:
\be\ba{ll}\medskip
\V I: \V u\in X\mapsto \V I\V u:=\Big( \V v+{\rm P}\,(\V\omega\times \V x),\ \mathbb I\cdot(\V\omega-\V a),\V z\Big)^\top\in X\,,\\ \medskip 
\V A:\V u\in {\sf D}(\V A):=[W^{2,2}(\calc)\cap W_0^{1,2}(\calc)\cap L^2_\sigma(\calc)]\oplus \real^3\oplus \real^3\subset X\mapsto \V A\V u:=\big(-\nu\,{\rm P}\,\Delta\V v\,,\ \V\omega,\V z)^\top\in X\,,\\ 
\V B: \V u\in X\mapsto \V B\V u:=\Big(2\,{\rm P}\,(-\lambda\,\V e_3\times\V v)\,,\ -\lambda\,\V e_3\times\mathbb I\cdot\V\omega-\lambda\,C\,\V\omega\times\V e_3+\lambda\,\V e_3\times\mathbb I\cdot\V a -\beta^2\V e_3\times\V z-\V\omega,\\ \medskip \hfill-\V \omega\times \V e_3-\lambda\V e_3\times \V z-\V z\Big)^\top\in X\,,\\
\V N: \V u\in  {\sf D}(\V A)\subset X\mapsto \V N(\V u):=\Big(-2 {\rm P}\,(\V\omega\times\V v)-{\rm P}\,(\V v\cdot\nabla\V v)\,,\ -\V\omega\times\mathbb I\cdot(\V\omega-\V a),-\V\omega\times\V z\Big)^\top\in X\,,
\ea
\eeq{1.6}
with $\mathrm P$ Helmholtz projection from $L^2(\calc)$ onto $L^2_\sigma(\calc)$. 
\par
From \cite[\S\,6.2.3]{KK}, we deduce the following.   
\Bl The operator $\V I$ is bounded, positive with a bounded inverse.
\EL{1}

In view of this lemma, we may set
\be
\A=\V I^{-1}\V A\,,\ \ \B=\V I^{-1}\V B\,,\ \ \L=\A+\B\,,\ \ \N=\V I^{-1}\V N\,, 
\eeq{newop}
where ${\sf D}(\L)={\sf D}(\N)\equiv {\sf D}(\V A)$, and deduce that
the system of equations \eqref{eq:perturbation} can be formally written as an  evolution equation in the Hilbert space $X$:
\be
\ode{\V u}t+\L\V u=\N(\V u)\,.
\eeq{1.7}
 
The stability  of the steady-state solution \eqref{s}  is then reduced to the  stability of the solution $\V u=\V0$ to \Eqref{1.7}. The study of the latter will be performed as a result of the general stability theory --  presented 
in next section that is founded upon suitable functional and spectral properties of the linear operator $\L$. 

\setcounter{equation}{0}
\section{\large Stability Properties for an Abstract Evolution Problem}\label{sec:abstract}
Objective of this section is to study the stability of the zero solution to a suitable evolution problem in a Banach space. The pecularity of this problem is that $0$ is an eigenvalue of the relevant linear operator, so that the classical ``linearization principle" (e.g. \cite[\S\,5.1]{henry}) does not apply.
We need, instead, a ``generalized linearization principle'' in the spirit of \cite[Theorem 2.1]{PSZ}. Here, we shall follow the approach of \cite{GMpendulum}, based on an operator fractional powers method, that appears to be more specific and direct for the type of fluid-structure
interaction problems considered in this paper; see, however, also \cite{MPS,MPS1}. In fact, the stability result stated in the following \theoref{1} is similar to its counterpart  in \cite[Theorem 1.1]{GMpendulum}, but obtained under slightly more general assumptions on the nonlinear operator and with a simpler argument. For completeness and reader's sake, in \theoref{UST} we also provide a short proof of the complementary instability result. \smallskip\par

In a (real) Banach space $X$, we consider the following evolution problem
\be
\ode{\V{u}}t+\L\V{u}=\N(\V{u})\,,\ \ \V{u}(0)\in X\,,
\eeq{1}
with $\L$ and $\N$ to be defined next. Let $\A:X\mapsto X$ be a linear, sectorial operator with compact inverse and {\color{black} $\mathsf{Re\;}\sigma(\A)\subset(0,\infty)$}. For $\alpha\in [0,1]$, set 
$$
X_{\alpha}=\big\{\V{u}\in X:\ \|\V{u}\|_\alpha:=\|\A^\alpha\V{u}\|<\infty\big\}\,;\ \ X_0\equiv X\,,\ \ \|\V{u}\|_0\equiv \|\V{u}\|\,.
$$  
It is well known that, for $\alpha>0$, $X_\alpha$ is a Banach space  compactly embedded in $X$,  e.g. \cite[Theorem 1.4.8]{henry}. Let $\B:X\mapsto X$ be a linear operator with ${\sf D}(\B)\supset {\sf D}(\A)$, and such that
\be
\|\B\V{u}\|\le c_1\,\|\V{u}\|_\alpha\,,\ \ \alpha\in [0,1)\,.
\eeq{0}
We then assume
\be 
\L=\A+\B\,,
\eeq{2}
with $D(\L)\equiv D(\A)$.
Since 
\be
\|\V{u}\|_\alpha\le c\,\|\V{u}\|^{1-\alpha}\|\A\V{u}\|^\alpha\,,
\eeq{2i}
and $\A^{-1}$ is compact, it follows  that $\B$ is $\A-$compact, which in turn implies  that  $\L$ is an unbounded Fredholm operator of index 0 \cite[Theorem 4.3]{GG}. Also, 
from \cite[Theorem 1.3.2]{henry} it follows, in particular, that $\L$ is sectorial. Finally, observing that by \Eqref{2i},  for any $\varepsilon>0$,
$$ 
\|\B\V{u}\|\le c(\varepsilon)\|\V{u}\|+\varepsilon\,\|\A\V{u}\|
$$
and that, by the properties of $\A$,
$$
\|(\lambda+\A)^{-1}\|\le c_2\lambda^{-1}\,,\ \ \mbox{all $\lambda>0$}
$$
it follows \cite[Theorem 3.17 at p. 214]{kato} that $\L$ has a compact resolvent and, therefore, a discrete spectrum. 
\par
On the operator $\L$, we further assume:
\tag{H1}
\be
{\rm dim}\,{\sf N}[\L]=m\ge1\,,\vspace*{-2mm}
\label{eq:H1}\ee
\tag{H2}
\be
{\sf N}[\L]\cap {\sf R}[\L]=\{\V{0}\}\,,
\label{eq:H2}\ee and
\tag{H3}
\be
\sigma(\L)\cap\{{\rm i}\,\real\}=\{0\}\,.
\label{eq:H3}
\ee\setcounter{equation}{4}
\renewcommand{\theequation}{\arabic{section}.\arabic{equation}}\smallskip\par
We then have the following.
\Bl The space $X$ admits the decomposition
\be
X={\sf N}[\L]\oplus{\sf R}[\L]\,.
\eeq{3}
Moreover, denoting by $\mathcal Q$ and $\mathcal P$ the spectral projections according to the spectral sets
$$\sigma_0(\L):=\{0\}\,,\ \
\sigma_1(\L):= \sigma(\L)\backslash\sigma_0(\L)\,,
$$
we have
\be
{\sf N}[\L]=\mathcal Q(X)\,,\ \ {\sf R}[\L]:=\mathcal P(X)\,.
\eeq{4}
Finally, \Eqref{3} completely reduces $\L$ into $\L=\L_0\oplus\L_1$
with
\be
\L_0:=\mathcal Q\L=\L\mathcal Q\,,\ \ \L_1:=\mathcal P\L=\L\mathcal P\,, 
\eeq{5}
and $\sigma(\L_0)\equiv\sigma_0(\L)$, $\sigma(\L_1)\equiv\sigma_1(\L)$.
\EL{1}
\begin{proof} Since $\L$ is Fredholm of index 0, from (\ref{eq:H1}) we get ${\rm codim}({\sf R}[\L])=m$. Thus, there exists at least one $S{\subset} X$ such that $X=S\oplus{\sf R}[\L]$, with $S\cap {\sf R}[\L]=\{\V{0}\}$.  However,  ${\rm dim}\,(S)={\rm dim}\,({\sf N}[\L])=m$ and (\ref{eq:H2}) holds, so that we may take $S={\sf N}[\L]$, which proves \Eqref{3}. The remaining properties stated in the lemma are then a consequence of \Eqref{3} and classical results on spectral theory (e.g., \cite[Proposition A.2.2]{Lun}, \cite[Theorems 5.7-A,B]{Tay})
\end{proof}\par
We now turn to the operator $\N$. We begin to assume
\tag{H4}
\be\ba{ll}\medskip
\|\N(\V{u}_1)-\N(\V{u}_2)\|\le c\,\|\V{u}_1-\V{u}_2\|_\alpha\,,\ \ \mbox{for all $\V{u}_1,\V{u}_2$ in a neighborhood of $\V{0}\in X_\alpha$}\,,\\
\|\N(\V{u})\|\le c\,\|\V{u}\|_\alpha^p\,,\ \ \mbox{for some $p>1$ and all $\V{u}$ in a neighborhood of $\V{0}\in X_\alpha$}\,.
\label{eq:H4}
\ea
\ee
Furthermore, we observe that, by \Eqref{3},  every $\V{u}\in X$ can be written as
$$
\V{u}=\V{u}^{(0)}+\V{u}^{(1)}\,,\ \ \V{u}^{(0)}\in {\sf N}[\L]\,,\ \V{u}^{(1)}\in {\sf R}[\L]\,.
$$
Thus, setting
$$
\M(\V{u}^{(0)},\V{u}^{(1)}):=\N(\V{u}^{(0)}+\V{u}^{(1)})
$$
we suppose there is a non-negative, continuous function $\epsilon=\epsilon(\rho)$ with $\epsilon(0)=0$ such that
\tag{H5}
\be
\|\M(\V{u}^{(0)},\V{u}^{(1)})\|\le \epsilon(\rho)\|\V{u}^{(1)}\|_\alpha\,,
\label{eq:H5}
\ee
whenever $\|\V{u}^{(0)}\|+\|\V{u}^{(1)}\|_\alpha \le C\, \rho$, some $C>0$.
\renewcommand{\theequation}{\arabic{section}.\arabic{equation}}\setcounter{equation}{7}
\smallskip\par
We are now in a position to prove the following stability result; see also \cite{GMpendulum}.
\Bt Suppose the operators $\L$, defined in \Eqref{2}, and $\N$ satisfy  hypotheses {\rm (\ref{eq:H1})--(\ref{eq:H4})$_1$ and (\ref{eq:H5})}. Then, if {\color{black} $\mathsf{Re\;}\sigma(\L_1)\subset (0,\infty)$}, we can find $\rho_0>0$ such that if
$$
\|\V{u}(0)\|_\alpha<\rho_0\,,
$$
there is a unique corresponding solution $\V{u}=\V{u}(t)$ to \Eqref{1} for all $t>0$, satisfying, for any $T>0$,
\be
\V{u}\in C([0, T);X_\alpha)\cap C((0,T); X_1)\cap C^1((0,T);X)\,.
\eeq{6}  
{\color{black} Moreover, the following two properties are satisfied: 
\begin{itemize}
\item[{\rm (a)}] The solution $\V{u}=\V{0}$ to \Eqref{1} is stable in $X_\alpha$, i.e., for any $\varepsilon>0$ there is $\delta>0$ such that
$$
\|\V{u}(0)\|_\alpha<\delta\ \ \Longrightarrow\ \ \sup_{t\ge 0}\|\V{u}(t)\|_\alpha <\varepsilon\,;
$$ 
\item[{\rm (b)}] There are $\eta,c,\kappa>0$ such that 
$$  
\|\V{u}(0)\|_\alpha<\eta\ \ \Longrightarrow\ \ \|\V{u}(t)-\bar{\V{u}}\|_\alpha\le c\,\|\V{u}^{(1)}(0)\|_\alpha\,{\rm 
e}^{-\kappa\,t}\,,\ \mbox{all $t>0$}\,,
$$
for some $\bar{\V{u}}\in {\sf N}[\L]$.
\end{itemize}}
\ET{1}
\begin{proof} Under the stated assumptions on $\A$, $\B$ and (\ref{eq:H4})$_1$, the
existence of a unique solution  $\V{u}$ to \Eqref{1} in some time interval $(0,t_\star)$ satisfying \Eqref{6} for each $T\in (0,t_\star)$ is guaranteed by classical results on semilinear evolution equations (e.g., \cite[p. 196--198]{Pazy}).
Moreover,  $(0,t^*)$ is maximal, in the sense that either $t^*=\infty$ or else
 $\lim_{t\to t^*}\|\V{u}(t)\|_\alpha=\infty$. 
We shall next show that, in fact,  only the former situation occurs for sufficiently ``small" initial data.
Applying  $\mathcal Q$ and $\mathcal P$  on both sides of \Eqref{1} 
 and taking into account \Eqref{5} we  get 
\be\ba{rl}\medskip
\ode{\V{u}^{(1)}}t+\L_1\V{u}^{(1)}=&\!\!\!\mathcal P\M(\V{u}^{(0)}, \V{u}^{(1)})\\
\ode{\V{u}^{(0)}}t =&\!\!\!\mathcal Q\M(\V{u}^{(0)}, \V{u}^{(1)})\,,
\ea
\eeq{1.20}
with $\V{u}^{(0)}=\mathcal Q\V{u}$, $\V{u}^{(1)}=\mathcal P\V{u}$.
Since the operator $\L$, being sectorial, is the generator of an analytic semigroup in $X$,  so is $\L_1$ in $X^{(1)}\equiv{\sf R}[\L]$.  Thus, for all $t\in[0,t_\star)$ from \Eqref{1.20}$_1$ we have
\be
\V{u}^{(1)}(t)={\rm e}^{-\L_1 t}\V{u}_0^{(1)}+\int_0^t{\rm e}^{-\L_1 (t-s)}[\mathcal P\M(\V{u}^{(0)}(s), \V{u}^{(1)}(s))]ds\,.
\eeq{1.22}
Also, by assumption and the spectral property of $\L$,  there is $\gamma>0$ such that 
{\color{black}\be
\mathsf{Re\;}\sigma(\L_1)\subset(\gamma, \infty)\,,  
\eeq{1.17}}
which implies that the fractional powers $\L_1^\alpha$, $\alpha\in (0,1)$, are well defined in $X^{(1)}$. Thus, setting
$$
\V{w}:={\rm e}^{bt}\L_1^\alpha\V{u}^{(1)}\,,\ \ 0<b<\gamma\,,
$$
from \Eqref{1.22} we get
\be
\V{w}(t)={\rm e}^{bt}{\rm e}^{-\L_1 t}\L_1^\alpha\V{u}_0^{(1)}+\int_0^t{\rm e}^{bt}\L_1^{\alpha} {\rm e}^{-\L_1 (t-s)}[\mathcal P\M(\V{u}^{(0)}(s), {\rm e}^{-bs}\L_1^{-\alpha}\V{w}(s)]ds\,.
\eeq{1.23}
From the local existence theory considered earlier on, we know that for any given $\rho>0$ there exists an interval of time $[0,\tau]$, $\tau<t_\star$, such that 
\be
\sup_{t\in [0,\tau)}\left(\|\V{w}(t)\|+\|\V{u}^{(0)}(t)\|\right)\le \rho\,,\ \tau<t_\star\,
\eeq{71}
provided 
$\|\V{u}(0)\|_\alpha<\eta$, for some $\eta>0$. 
Let us show that  $\eta$ and $\rho$ can be chosen sufficiently small so that \Eqref{71} holds also with $\tau=t_\star$, thus implying, in particular, that the solution $\V{u}=\V{u}(t)$ to \Eqref{1} exists for all times $t>0$. In fact, suppose, 
by contradiction, that there is $\tau_0<t_\star$ such that
\be
\|\V{w}(t)\|+\|\V{u}^{(0)}(t)\|<\rho\,,\ \ t\in [0,\tau_0)\ \ \mbox{and}\ \ \|\V{w}(\tau_0)\|+\|\V{u}^{(0)}(\tau_0)\|=\rho\,.
\eeq{72}
In view of the stated properties of $\L_1$ it results (e.g. \cite[Theorem 1.4.6]{henry}), 
\be
\|\L_1^{-\alpha}\V{w}\|_\alpha\le c_1\,\|\V{w}\|\,,\ \  \mbox{for all $\V{w}\in X^{(1)}$}\,,
\eeq{1.30}
Moreover, we recall that in $X^{(1)}$ it is 
\be\|\L_1^{\alpha}{\rm e}^{-\L_1t}\|\le C_\alpha t^{-\alpha}{\rm e}^{-\gamma t}\,, 
\eeq{df0} 
and observe that from \Eqref{0} and 
 \cite[Theorem 1.4.6]{henry} 
\be
\|\L_1^\alpha\V{u}_0^{(1)}\|\le c_2\,\|\V{u}_0^{(1)}\|_\alpha\,.
\eeq{df1}
Thus, noticing that
$$
\int_0^t\frac{{\rm e}^{-(\gamma-b)(t-s)}}{(t-s)^\alpha}\,ds\le \int_0^\infty\frac{{\rm e}^{-(\gamma-b)t}}{t^\alpha}\,dt<\infty\,,
$$
from  \Eqref{1.23}, \Eqref{72}--\Eqref{df1}, and (\ref{eq:H5}) we show
\be
\|\V{w}(t)\|\equiv {\rm e}^{bt}\|\V{u}^{(1)}(t)\|_\alpha\le C_0\eta + \epsilon(\rho)\, \rho\,,\ \mbox{for all $t\in [0,\tau_0]$}\,.
\eeq{1.31}
On the other hand, \Eqref{1.20}$_2$,  with the help of \Eqref{72}, (\ref{eq:H5}) and \Eqref{1.31}, furnishes
\be\ba{ll}\medskip
\|\V{u}^{(0)}(t)\|\le &\!\!\!\|\V{u}^{(0)}(0)\|+\epsilon(\rho)\Int0t \,\|\V{u}^{(1)}(s)\|_\alpha\,ds\le \|\V{u}^{(0)}(0)\|+\epsilon(\rho)\Int0t \,{\rm e}^{-b\,s}\|\V{w}(s)\|\,ds \\&\!\!\!\!\le c_3\eta +\epsilon(\rho)\,\rho\,,\ \ t\in [0,\tau_0]\,. \ea
\eeq{1.32} 
Combining \Eqref{1.31} and \Eqref{1.32}, and choosing $c_3\eta/\rho+ \epsilon(\rho)<1/4$ we  conclude in particular
$$
\|\V{w}(\tau_0)\|+\|\V{u}^{(0)}(\tau_0)\|\le \rho/2
$$
contradicting \Eqref{72}. As a result, by what we observed early on, we may take $t_*=\infty$ in \Eqref{71} and conclude as well
\be
\sup_{t\in [0,\infty)}\left(\|\V{w}(t)\|+\|\V{u}^{(0)}(t)\|\right)\le \rho\,,
\eeq{711}
proving, as a byproduct, the desired global existence property.
Now, employing in \Eqref{1.23}, the inequalities  \Eqref{711},   \Eqref{1.30} along with the assumption \Eqref{H5}, we easily deduce, for $\rho$ small enough, 
\be
\|\V{w}(t)\|\le c_4\,\|\V{u}^{(1)}(0)\|_\alpha\,,\ \ \mbox{all $t>0$}\,,
\eeq{sfc}
namely,
\be
\|\V{u}^{(1)}(t)\|_\alpha\le c_4 \,{\rm e}^{-b t} \|\V{u}^{(1)}(0)\|_\alpha\,,\ \ \mbox{all $t>0$}\,.  
\eeq{1.35}
Also, using \Eqref{sfc} into \Eqref{1.32}, we infer
\be
\|\V{u}^{(0)}(t)\|\le c_5\,\|\V{u}(0)\|_\alpha\,,\ \ \mbox{all $t>0$}\,.
\eeq{sfc1}
Therefore, from \Eqref{1.35} and \Eqref{sfc1} we recover the stability property stated in (a).
Moreover, integrating  \Eqref{1.20}$_2$ between arbitrary $t_1,t_2>0$ using \Eqref{1.35} and reasoning in a way similar to what we did to obtain \Eqref{1.32} we get
\be
\|\V{u}^{(0)}(t_1)-\V{u}^{(0)}(t_2)\|\le c_{6}\int_{t_1}^{t_2}\|\V{u}^{(1)}(s)\|_\alpha ds\le c_{7}\,\|\V{u}^{(1)}(0)\|_\alpha\int_{t_1}^{t_2}{\rm e}^{-b s}ds\,, 
\eeq{1.36}
from which we deduce that there exists $\bar{\V{u}}\in {\sf N}[\L]$ such that
$$
\lim_{t\to\infty}\|\V{u}^{(0)}(t)-\bar{\V{u}}\|=0\,.
$$
Employing this information into \Eqref{1.36} in the limit $t_2\to\infty$, and with $t_1=t$ we show
$$
\|\V{u}^{(0)}(t)-\bar{\V{u}}\|_\alpha\le c_{8}\,\|\V{u}^{(0)}(t)-\bar{\V{u}}\|\le c_{9}\|\V{u}^{(1)}(0)\|_\alpha\,{\rm e}^{-b t},
$$
which, once combined with \Eqref{1.35}, proves the exponential rate of decay stated in (b).   The proof of the theorem is thus concluded.
\end{proof}\par
We next show the following instability result.\footnote{The results stated in \theoref{UST} could also be obtained by means of \cite[Theorem 5.1.5]{henry}. However, for completeness, we present a simple proof by a different method.}
\Bt Let $\L$ be the operator defined in \Eqref{2}, and suppose that $\N$ satisfies hypotheses {\rm (\ref{eq:H1})--(\ref{eq:H4}}) Then, if ${\sf Re}\,\sigma(\L)\cap (-\infty,0)\neq\emptyset$ the solution $\V{u}=\V{0}$ to \Eqref{1} is unstable in $X_\alpha$. Precisely, there is $\varepsilon_0>0$ with the following property: for any given $\delta\in (0,1]$ there is a solution $\V{u}(t)$ to \Eqref{1} in the class \Eqref{6} with $\|\V{u}(0)\|_\alpha\le \delta$ such that
$$
\sup_{t\in (0,t^*)}\|\V{u}(t)\|_\alpha\ge \varepsilon_0\,,$$ where $(0,t^*)$ is the maximal interval of existence.
\ET{UST}
\begin{proof} We follow the ideas of \cite[Theorem 2.3 in Chapter VII]{DK}. From \Eqref{1} we deduce
\be
\V{u}(t)={\rm e}^{-\L\,t}\V{u}(0)+\int_0^t{\rm e}^{-\L(t-s)}\N(\V{u}(s))\,ds:=\V{u}_1(t)+\V{u}_2(t)\,.
\eeq{p0}
Let $\lambda_0$ be the eigenvalue of $\L$ such that 
$$
-a_0:={\sf Re}\,\lambda_0=\min_{\lambda\in \sigma(\L)}{\sf Re}\,\lambda\,.
$$
By assumption, $a_0>0$ and we write $\lambda_0=-a_0+{\rm i}\,b_0$,  $b_0\in\real$. Let $\V{\varphi}\in {\sf D}(\L)$ $(\equiv{\sf D}(\A))$ be the eigenvector corresponding to $\lambda_0$, normalized with $\|\V{\varphi}\|_\alpha=1$, and choose $\V{u}(0)=\delta\,{\sf Re}\,\V{\varphi}$. Since ${\rm e}^{-\L\,t}\V{u}(0)=\delta\,{\rm e}^{(a_0+{\rm i}\,b_0)\,t}\V{\varphi}$ we get, therefore,
\be
\|\V{u}_1(t)\|_\alpha\le\delta\,{\rm e}^{a_0\,t}\,.
\eeq{p1}
Next, fix $R\in (1,2)$ and denote by $\tau_0$ $(<t^*)$ the largest positive number for which
\be 
\|\V{u}(t)\|_\alpha\le \delta\,R\,{\rm e}^{a_0\,t}\,,\ \ t\in[0,\tau_0]\,.
\eeq{p2}
Let $\L_\zeta:=\L+\zeta$, with $\zeta>a_0$. Clearly, {\color{black} $\mathsf{Re\;}\sigma(\L_\zeta)\subset(0,\infty)$}, so that $\L_\zeta^{-\alpha}$ is well defined. We thus get
$$
\A^\alpha {\rm e}^{-\L(t-s)}\N(\V{u}(s))={\rm e}^{\zeta(t-s)}\A^\alpha\L_\zeta^{-\alpha}
\L_\zeta^{\alpha}{\rm e}^{-\L_\zeta(t-s)}\N(\V{u}(s))
$$
which, in turn, after observing that $\A^\alpha\L_\zeta^{-\alpha}$ is bounded in $X$ \cite[Theorem 1.4.6]{henry}, and taking into account (\ref{eq:H4})$_2$, implies
\be
\|\A^\alpha {\rm e}^{-\L(t-s)}\N(\V{u}(s))\|\le C\,{\rm e}^{\zeta(t-s)}\,\|\L_\zeta^{\alpha}{\rm e}^{-\L_\zeta(t-s)}\|\,\|\V{u}(s)\|_\alpha^p\,.
\eeq{p3}
Therefore, since
\be
\|\L_\zeta^{\alpha}{\rm e}^{-\L_\zeta(t-s)}\|\le\, c_\alpha\, \frac{{\rm e}^{(a_0-\zeta)(t-s)}}{(t-s)^\alpha}\,,
\eeq{p4}
from \Eqref{p0}, and \Eqref{p2}--\Eqref{p4} we show
\be\ba{rl}\medskip
\|\V{u}_2(t)\|_\alpha&\!\!\!\le c_1\,\Int{0}t{\rm e}^{\zeta(t-s)}\Frac{{\rm e}^{(a_0-\zeta)(t-s)}}{(t-s)^\alpha}\|\V{u}(s)\|_\alpha^p\,ds= c_1\,\Int{0}t\Frac{{\rm e}^{a_0(t-s)}}{(t-s)^\alpha}\|\V{u}(s)\|_\alpha^p\,ds
\\
&\le
\,c_1\,\delta^pR^p{\rm e}^{p a_0 t}\Int0t\Frac{{\rm e}^{a_0(1-p)\rho}}{\rho^\alpha}\,d\rho\le c_0 \,\delta^pR^p{\rm e}^{p\,a_0\,t}\,,\ \ t\in [0,\tau_0]\,,
\ea
\eeq{p5}
where $c_0=c_0(\alpha,a_0)$. Define the positive number $\tau$ through the relation
\be
1+c_0 \,\delta^{p-1}R^p{\rm e}^{(p-1)\,a_0\,\tau}=R\,.
\eeq{p6}
It is clear that $\tau\le \tau_0$, because otherwise from \Eqref{p0}, \Eqref{p1}, \Eqref{p5} and \Eqref{p6} we would contradict the definition of $\tau_0$ for which \Eqref{p2} holds. However, for this $\tau$, from \Eqref{p0}, \Eqref{p1},  \Eqref{p5} and \Eqref{p6} it follows that
$$
\|\V{u}(\tau)\|_\alpha\ge \delta\,{\rm e}^{a_0\tau}\big(1-c_0\delta^{p-1}R^p{\rm e}^{(p-1)a_0\tau}\big)=\left(\Frac{R-1}{c_0\,R^p}\right)^{\frac1{p-1}}(2-R):=\varepsilon_0\,.
$$
The proof of the theorem is completed.\end{proof}

\setcounter{equation}{0}
\section{\large Preliminary Properties of the Operator $\L$}
{\color{black}The main purpose of this section is to find conditions under which the operator $\L$ defined in  \Eqref{newop}, \Eqref{1.6} satisfies the assumptions \Eqref{H1}--\Eqref{H3} stated in the previous section. 
In this regard we begin to observe that, by the well known properties of the Stokes operator  
\begin{equation}\label{eq:stokes}
\V A_0:=-\nu\,{\rm P}\Delta
\end{equation}
with domain ${\sf D}(\V A_0):=W^{2,2}(\mathscr C)\cap \mathcal D_0^{1,2}(\mathscr C)$ and range  $L^2_\sigma(\mathscr C)$,
it follows that ${\V A}$ has a compact inverse and, therefore, a discrete spectrum which, in addition, lies on the positive real axis. Since $\V I^{-1}$ is symmetric (and bounded), the operator ${\A}$ enjoys the same properties as ${\V A}$. Furthermore,  $\V B$ is bounded and therefore $\B$ satisfies \Eqref{0} with $\alpha=0$. We shall now check the validity of the other assumptions. 

\smallskip\par 
We start with the following  result which, in particular, gives sufficient conditions in order that $\L$ satisfies \Eqref{H1}.
\Bp Let $\bar{\sf s}$ be the steady-state solution \eqref{s} and let $\L=\L(\bar{\V s})$ be the linearization around $\bar{\sf s}$. Then,  setting 
\be
K_A:=\frac{\lambda^2}{\beta^2}(C-A)\,,\ \ K_B:=\frac{\lambda^2}{\beta^2}(C-B)\,,
\eeq{K} 
we have 
\be
{\sf N}[\L]=\{\V v\equiv\V0;\, \V\omega=r\,\V e_3;\, \V z=z\,\V e_3;\,\ (r,z)\in\real^2\}\,,\ \ {\rm dim}\,{\sf N}[\L]=2\,,
\eeq{C1}
provided $K_A,K_B\neq 1$.
\EP{3.1}
\begin{proof} From \lemmref{1} and  \Eqref{1.6}$_{2,3}$, it follows that the equation $\L\V u=\V0$ is equivalent to the following system of equations
\begin{equation}
\begin{array}{ll}\medskip
\left.\begin{array}{ll}\smallskip
\nu\Delta \V v-\nabla p=-2\lambda\V e_3\times \V v
\\
\diver \V v=0
\end{array}\right\}\ \ \text{in }{\mathscr C}\,,
\\
\hspace*{3mm}\V v=\V 0\quad \text{on }\partial \mathscr C\,,
\medskip
\\\ba{ll} -\lambda\V e_3\times\T I\cdot(\V\omega-\V a)+\lambda\, C\,\V e_3\times \V \omega=\beta^2\V e_3\times \V z \,,
\medskip \\
-\V \omega\times \V e_3-\lambda\,\V e_3\times \V z=\V 0\ea\,.
\end{array}\eeq{3.1}
By dot-multiplying both sides of \Eqref{3.1}$_1$  by $\V v$,  integrating by parts over $\mathscr  C$ and using  \Eqref{3.1}$_{2,3}$, we deduce $\V v\equiv\V0$. As a result, the elements of ${\sf N}[\L]$ must have $\V v\equiv\V0$, while $\V\omega$ and $\V z$ solve the following equations
\be 
\ba{ll} -\lambda\,\V e_3\times \T I\cdot\V\omega+\V e_3\times \big(\lambda\, C\,\V\omega-\beta^2\V z\big)=\V 0 \,,
\medskip \\
\V \omega=\lambda\,\V z+\mu\,\V e_3\,,\ \ \mu\in\R\,.\ea
\eeq{3.2}
Replacing \Eqref{3.2}$_2$ into \Eqref{3.2}$_1$ and recalling that $\T I={\rm diag}\,(A,B,C)$ we thus get
$$
(1-K_B)z_2\V e_1+(K_A-1)z_1\V e_2=\V 0\,,
$$
which, in turn, under the given assumptions furnishes $\V z=z\V e_3$, $z\in\R$. The desired property is then a consequence of the latter and of \Eqref{3.2}$_2$.
\end{proof}
We now pass to the investigation of the validity of the assumption \Eqref{H2}. 
{\color{black}
\Bp Let the assumptions of  \propref{3.1} be satisfied. Then  ${\sf N}[\L]\cap {\sf R}[\L]=\{\V 0\}$.
\EP{3.2}
\begin{proof} From \Eqref{1.6}, \Eqref{newop} and \propref{3.1}, we deduce that the condition ${\sf N}[\L]\cap {\sf R}[\L]=\{\V 0\}$ is equivalent to show that, if $(\V v_0\equiv 0,\V\omega_0=\omega_0\V e_3,\V z_0=z_0\V e_3)\in {\sf N}[\L]$, and 
\be\ba{ll}\medskip
\left.\ba{ll}\medskip-\nu\Delta\V v-2\lambda\V e_3\times\V v +\nabla p=\omega_0\V e_3\times\V x\\
\diver \V v=0\ea\right\}\ \mbox{in $\mathscr C$}\,,\\
\hspace*{2.5mm}\V v=\V 0\ \ \mbox{at $\partial\mathscr C$}
\medskip\\
\ba{ll}\medskip
-\lambda\V e_3\times \T I\cdot (\V \omega-\V a)-\lambda C\V\omega\times \V e_3-\beta^2\V e_3\times \V z=C\omega_0\V e_3\\ 
(\V\omega-\lambda\V z)\times\V e_3=-z_0\V e_3
\ea
\ea
\eeq{3.22}
for some $(\V v,\V\omega,\V z)\in {\sf D}(\V A)$, then necessarily $\V\omega_0=\V z_0=\V 0$. However, the latter follows at once by dot-multiplying both sides of 
 \Eqref{3.22}$_{5,6}$ by $\V e_3$. \end{proof}

}
In order to investigate further properties of the operator $\L$, we need the following.
\Bl Let
\be
\mathscr G=\mathscr G(\V u):=\|\V v\|_2^2-\V a\cdot\T I\cdot\V a+\V\omega_*\cdot\T I\cdot\V\omega_*+\delta\, |\V z|^2-2\lambda\, \V z\cdot\T I\cdot\V\omega_*\,,
\eeq{Gf}
where $\V\omega_*:=\V\omega-\V a$ (with $\V a$ given in \eqref{eq:notation}) and 
\be\delta= \lambda^2C-{\beta^2}\,.\eeq{3.33}
Then 
\be
\half\ode{\mathscr G}t+\nu\|\nabla\V v\|_2^2=0
\eeq{GP}
along the solutions to the linear problem
\be
\ode{\V u}{t}+\L\V u=\V0\,,\ \ \V u(0)\in X\,.
\eeq{3.34}
\EL{3.3}
\begin{proof} From \Eqref{1.7} we deduce that the evolution equation in \Eqref{3.34} is equivalent to the following set of equations:
\be
\begin{array}{ll}\medskip
\left.\begin{array}{ll}\smallskip
\V v_t+(\dot{\V\omega}_*+\dot{\V a})\times \V x-2\lambda{\V\e_3}\times \V v=\nu\Delta \V v-\nabla p
\\
\diver \V v=0
\end{array}\right\}\ \ \text{on }{\mathscr C}\times (0,\infty)\,,\medskip
\\\left.\ba{ll} \T I\cdot \dot{\V \omega}_*
-\lambda{\V e_3}\times\T I\cdot\V\omega_*-\lambda C(\V\omega_*+\V a)\times {\V e_3}=\beta^2\V e_3\times \V z 
\medskip \\
\V{\dot z}-(\V \omega_*+\V a)\times \V e_3-\lambda {\V e_3}\times \V z=\V 0\ea\right\}\ \mbox{in $(0,\infty)$}\,,\medskip
\\
\hspace*{2mm} \V v=\V 0\quad \text{on }\partial \mathscr C\times(0,\infty)\,.
\end{array}
\eeq{3.35}
Due to the analyticity of the semigroup generated by $\L$, solutions to \Eqref{3.35} with initial data in $X$ are smooth. We dot-multiply both sides of \Eqref{3.35}$_1$ by $\V v$ and integrate by parts over $\mathscr C$ to get
\be
\half\left(\ode{}t\|\V v\|_2^2-\V a\cdot\T I\cdot\V a\right)-\dot{\V\omega}_*\cdot\T I\cdot\V a+\nu\|\nabla\V v\|_2=0\,.
\eeq{3.36}
We next dot-multiply {\color{black}both sides of \Eqref{3.35}$_3$ by $\V z$ and both sides of \Eqref{3.35}$_4$ by }$\T I\cdot\V\omega_*$, and sum the resulting equations side-by-side. We thus infer
\be 
\ode{}t(\V z\cdot\T I\cdot\V\omega_*)-(\V\omega_*-\V a)\times\V e_3\cdot(\T I\cdot\V\omega_*)-\lambda C(\V\omega_*+\V a)\times\V e_3\cdot\V z=0\,.
\eeq{3.37}
By dot-multiplying both sides of \Eqref{3.35}$_3$ by $\V\omega_*+\V a$ we show
\be
\half\ode{}t(\V\omega_*\cdot\T I\cdot\V\omega_*)-\lambda\V e_3\times(\T I\cdot\V\omega_*)\cdot(\V\omega_*+\V a)+\V a\cdot\T I\cdot\dot{\V\omega}_*-\beta^2\V e_3\times\V z\cdot(\V\omega_*+\V a)=0\,.
\eeq{3.38}
Finally, dot-multiplying both sides of \Eqref{3.35}$_4$ by $\V z$ we obtain
\be
\half \ode{}t|\V z|^2-(\V\omega_*-\V a)\cdot\V e_3\times\V z=0\,.
\eeq{3.39}
If we form the linear combination $\Eqref{3.36}+\Eqref{3.38}+\delta\,\Eqref{3.39}-\lambda\,\Eqref{3.37}$ and use \Eqref{3.33}, we deduce 
$$
\half\ode{\mathscr G}t+\nu\|\nabla\V v\|_2^2=(\V\omega_*+\V a)\cdot\left[(\beta^2\V e_3+\delta\,\V e_3-\lambda^2\,C\V e_3)\times\V z\right]=0\,,
$$
which completes the proof. \end{proof}

A first important consequence of this lemma is provided by the following result that furnishes sufficient conditions for the validity of \Eqref{H3}. 
\Bp  $\sigma(\L)\cap \{{\rm i}\,\real\}=\{0\}$.
\EP{3.4}
\begin{proof} Contradicting the stated property means that the system \Eqref{3.35} has at least one time-periodic solution $(\V v,\V\omega_*,\V z)$  of period $T>0$ (say) such that 
\be
\int_0^T\V v(x,t)\,\d t=\int_0^T\V \omega_*(t)\,\d t=\int_0^T\V z(t)\,\d t=\V0\,.
\eeq{3.42}
Integrating \Eqref{GP} from $0$ to $T$ and using the periodicity we deduce 
$$
\int_0^T\|\nabla\V v\|_2^2\,\d t=0
$$
which, in turn, by \Eqref{3.35}$_5$ furnishes $\V v\equiv \V 0$. Then, by \Eqref{3.35}$_1$ it is $\dot{\V\omega}_*=-\nabla p$ and by operating with ${\rm curl}$ on both sides we get $\V\omega_*=\textrm{\bf const}$, implying, by \Eqref{3.42}, $\V\omega_*=\V 0$. The latter in conjunction with \Eqref{3.35}$_3$ delivers $\V z=z_3\V e_3$ that, once combined with \Eqref{3.35}$_4$, allows us to deduce $z_3=\textrm{const.}$ Thus, by \Eqref{3.42} we find also that $\V z=\V0$, and the proof of the proposition is completed.\end{proof}

We conclude this section by collecting in the following theorem some relevant consequences of \propref{3.1}, \propref{3.2} and \propref{3.4}.
\Bt Suppose  the assumptions  of \propref{3.1} are satisfied. Then, the linear operator $\L=\L(\bar{\sf s})$ meets all conditions \Eqref{H1}--\Eqref{H3}. 
\ET{3.5}

In the next section we shall give detailed information on the spectrum of $\L$.
\setcounter{equation}{0}
\section{\large On the Location of the Spectrum  of the Operator $\L$}
We now turn to the study of the location of the eigenvalues of the operator $\L$. As a matter of fact, 
 according to the decomposition proved in \lemmref{1}, this amounts to investigate the same property for the restriction, $\L_1$, of $\L$ to the subspace ${\sf R}[\L]$.

To this end, we propose two preparatory results collected in the form of as many lemmas.
\Bl  Suppose the assumptions of \propref{3.1} hold, and that the number $\delta$, defined in \Eqref{3.33}, is such that
\be
\delta>\lambda^2\,\max\{A,B\}\,.
\eeq{4.1} 
Then, all solutions $\V u$ to problem \Eqref{3.34} satisfy the estimate
\be
\sup_{t\ge 0}\|\V u(t)\|\le \kappa
\eeq{4.2}
for some $\kappa=\kappa(\V u(0), A,B,C)>0$.
\EL{4.1}
\begin{proof} Setting $\V\omega_{*}=(\zeta_1,\zeta_2,\zeta_3)$, we define
\be\ba{ll}\medskip
\mathscr Q_1=\mathscr Q_1(\zeta_1,z_1):=A\,\zeta_1^2+\delta\,z_1^2-2\lambda\,A\,\zeta_1z_1\,;\\ \medskip \mathscr Q_2=\mathscr Q_2(\zeta_2,z_2):=B\,\zeta_2^2+\delta\,z_2^2-2\lambda\,B\,\zeta_2z_2\,;\\ \mathscr Q_3=\mathscr Q_3(\zeta_3,z_3):=C\,\zeta_3^2+\delta\,z_3^2-2\lambda\,C\,\zeta_3z_3\,,\ea
\eeq{Q}
so that the functional $\mathscr G$ in \Eqref{Gf} can be written as
\be
\mathscr G=\|\V v\|_2^2-\V a\cdot\T I\cdot\V a+\mathscr Q_1+\mathscr Q_2+\mathscr Q_3\,.
\eeq{Gg}
We next observe that by \cite[\S\S\, 7.2.2, 7.2.3 ]{KK}, there is $\kappa_0\in(0,1)$ such that
\be
\kappa_0\|\V v\|_2^2\le \mathcal E_F:=\|\V v\|_2^2-\V a\cdot\T I\cdot\V a\le \|\V v\|_2^2\,. 
\eeq{4.3}
Now, since $\delta>\lambda^2\max\{A,B,C\}$,  all quadratic form $\mathscr Q_i$, $i=1,2,3$, are positive definite. As a consequence, integrating both sides of \Eqref{GP} from $0$ to $t$ and using this information along with \Eqref{4.3} leads to \Eqref{4.2}. \end{proof}
\Bl
Let $\V u$ be a solution to \Eqref{3.34} with initial data $\V u(0)=\V u_0$, such that 
\be
\sup_{t\ge 0}\|\V u(t)\|\le \kappa\,, 
\eeq{4.5}
for some constant $\kappa>0$. Then, the corresponding $\omega$-limit set, $\Omega(\V u_0)$ is not empty and satisfies
$$
\Omega(\V u_0)\subseteq {\sf N}[\L]\,.
$$
\EL{4.2}
\begin{proof} We recall that {\Eqref{3.34} is equivalent \Eqref{3.35}} endowed with the appropriate initial data. Thus, from \Eqref{4.5} and \Eqref{3.35}$_3$, it also follows that $|\V{\dot \omega}_\star|$ is uniformly bounded in time. Therefore \Eqref{3.36}, together with Poincar\'e inequality and \Eqref{4.3}, implies  
\begin{equation}\label{eq:energy_v}
\frac{\d}{\d t}\mathcal E_F(\V v)+c_1\mathcal E_F(\V v)\le c_2\norm{\V v}_2.
\end{equation}
Furthermore, {by \Eqref{GP}, \Eqref{4.5} and Poincar\'e inequality}, it follows that $\V v\in L^2(0,\infty;L^2_\sigma(\mathscr C))$. The latter, combined with \eqref{eq:energy_v} and \Eqref{4.3}, allows us to conclude that (see  \cite[Lemma 2.3.4]{Ma2})
\[
\lim_{t\to \infty}\norm{\V v(t)}_2=0.
\]
From this and  {again \Eqref{4.5}}, it follows that the $\omega$-limit set $\Omega(\V u_0)$ of the dynamical system generated by \Eqref{3.34} must be connected, compact and invariant. Moreover, 
\[
\lim_{t\to \infty}\mathsf{dist} (\V u,\Omega(\V u_0))=0. 
\]
The dynamics on the $\omega$-limit set is then characterized by $\V v\equiv\V0$, and $\V\omega \equiv \V\omega_*$, and $\V z$ satisfying
$$
\begin{array}{ll}\medskip
\dot{\V\omega}\times \V x=-\nabla p
\\
\T I\cdot{\dot{\V\omega}}-\lambda {\V e_3}\times\T I\cdot{\V\omega}-\lambda\, C\V\omega\times \V e_3=\beta^2\V e_3\times \V z \,,\medskip
\\
\dot{\V z}-\V \omega\times \V e_3-\lambda\, C\,{\V e_3}\times \V z=\V 0\,.
\end{array}
$$
By taking the curl of both sides of the first equation, it immediately follows  $\V{\dot \omega}=\V 0$, which, in turn, by the second equation, furnishes $\dot{z}_i=0$, $i=1,2$. Using this information in the third equation dot-multiplied by $\V e_3$ we get also $\dot{z}_3=0$. As a consequence,  the previous system reduces to 
$$
\begin{array}{ll}\medskip
-\lambda\,\V e_3\times\T I\cdot{\V\omega}+\V e_3\times(\lambda \,C\,\V\omega-\beta^2 \V z)=\V 0 \,,\medskip
\\
-\V \omega\times \V e_3-\lambda \,\V e_3 \times \V z=\V 0\,,
\end{array}
$$
which coincides with \Eqref{3.2}. As a result, proceeding as in the proof of \propref{3.1}, we show $\Omega(\V u_0)\subseteq{\sf N}[\L]$.\end{proof}

We are now in a position to prove the main result of this section. 
\Bt
Set $M=\max\{A,B\}$. The following properties hold. 
\begin{itemize}
\item[{\rm (i)}] If $C>M$ and 
\begin{equation}\label{eq:lin_stability2}
\lambda^2>\frac{\beta^2}{C-M}\,,
\end{equation}
then $\mathsf{Re\;}\sigma(\L_1)\subset(0,\infty)$. 
\item[{\rm (ii)}] If $C>M$ and 
\begin{equation}\label{eq:lin_instability2}
\lambda^2<\frac{\beta^2}{C-M}\,
\end{equation}
then $\mathsf{Re\;}\sigma(\L_1)\cap(-\infty,0)\ne\emptyset$. 
\item[{\rm (iii)}]  If $C< M$, then $\mathsf{Re\;}\sigma(\L_1)\cap(-\infty,0)\ne\emptyset$.
\end{itemize}

\ET{4.3}
\begin{proof} Our  strategy in showing the theorem goes as follows. To prove $\mathsf{Re\;}\sigma(\L_1)\subset(0,\infty)$, it is (necessary and) sufficient to show that all solutions to \Eqref{3.35} are uniformly bounded in time. To this purpose, we shall use \lemmref{4.1}. Conversely, to prove $\mathsf{Re\;}\sigma(\L_1)\cap(-\infty,0)\ne\emptyset$, it suffices to show that there exists at least one unbounded solution, and this will be achieved by a contradiction argument that exploits \lemmref{4.2}. With this in mind, we now proceed to the proof of the theorem.
\par\noindent\begin{itemize}
\item[{\rm (i)}] Since  $C>A,\,B$,  \lemmref{4.1}  shows that solutions are bounded if \eqref{eq:lin_stability2} holds. 
\item[{\rm (ii)}] Suppose, by contradiction,  $\mathsf{Re} \sigma(\L_1)\subset (0,\infty)$, then every solution to \Eqref{3.34} (or, equivalently, \Eqref{3.35}) must be bounded and so, by \lemmref{4.2}, $\Omega(\V u_0)\subseteq{\sf N}[\L]$. By assumption and \propref{3.1} it thus follows  
\be\Omega(\V u_0)=\{\V v\equiv\V 0\,;\, \V\omega=r\,\V e_3\,;\ \V z=z\,\V e_3\,;\ \mbox{some} \,(r,z)\in\real^2\}\,.
\eeq{om}  
Thus, denoting by $\{t_n\}$ an unbounded sequence of times, integrating both sides of \Eqref{GP} over $(0,t_n)$ and taking the limit as $n\to \infty$, we obtain  with {$\V\omega_0=(p_0,q_0,r_0)$ and $\V z_0=(z_{01},z_{02},z_{03})$}
$$
0\le\mathscr Q_3(r,z)\le \mathcal E_F(\V v(0))+ \mathscr Q_1(p_0,z_{01})+\mathscr  Q_2(p_0,z_{02})+\mathscr Q_3(r_0,z_{03})
\,.
$$
where we used \Eqref{Q} and \Eqref{Gg}.
Choosing $\V v(0)\equiv\V 0$,  $r_0=z_{03}=0$, $p_0=\lambda\,z_{01}$, $q_0=\lambda\,z_{02}$, $z_{0i}\neq 0$, $i=1,2$, the previous inequality along with \eqref{eq:lin_instability2} entails  
\be
0\le -\lambda^2z_{01}^2[A-C+\frac{\beta^2}{\lambda^2}]-\lambda^2z_{02}^2[B-C+\frac{\beta^2}{\lambda^2}]<0\,, 
\eeq{4.14}
which gives a contradiction.
\item[{\rm (iii)}] By contradiction, 
assume, instead, $\mathsf{Re} \sigma(\L_1)\subset (0,\infty)$. From \Eqref{3.35}$_4$ it follows that $z_3=\textrm{const.}$, so that \Eqref{GP} can also be equivalently written as
\be
\ode{\mathscr G_1}{t}+\nu\|\nabla\V v\|_2^2=0
\eeq{VF}
where
$$
\mathscr G_1:=\|\V v\|_2^2-\V a\cdot\T I\cdot\V a+\mathscr Q_1+\mathscr Q_2+\mathscr Q_3^*\,;\ \ \ \mathscr Q_3^*:=C(\zeta_3-\lambda\,z_3)^2\,.
$$
Thus, integrating \Eqref{VF} from $0$ to $t$ and
arguing exactly like in {\rm (ii)}, we arrive again at \Eqref{4.14}. Now, if $C<A,B$, we get an immediate contradiction. If $A<C<B$ we take $z_{01}=0$, thus getting a contradiction again. The case $B<C<A$ is treated analogously, which complete the proof of the theorem. 
\end{itemize}
\end{proof}
\setcounter{equation}{0}
\section{\large Properties of the  Operator $\N$}
We will now prove that the nonlinear operator $\N$ defined in \Eqref{newop}, \Eqref{1.6} satisfies the assumptions in \Eqref{H4}, \Eqref{H5}. The latter, together with the results proved in the preceding two sections, will thus allow us to use \theoref{1} and \theoref{UST}, and provide a complete characterization of the stability and asymptotic stability of the steady motions for \eqref{eq:motion}; see Section 6. 

Let us use the canonical decomposition $X_\alpha=Y_\alpha\times \R^3\times\R^3$, where 
\[
Y_\alpha:=\{\V v\in L^2_\sigma(\mathscr C):\; \norm{\V A_0^\alpha\V v}_2<\infty\}
\]
and the operator $\V A_0$ is the Stokes operator introduced in \eqref{eq:stokes}. Then, 
\begin{equation}\label{eq:interpolation}
Y_\alpha\subset H^{2\alpha}(\mathscr C):=[L^2(\mathscr C), W^{2,2}(\mathscr C)]_\alpha
\end{equation}
where $[\cdot,\cdot]_\alpha$ denotes the complex interpolation. The nonlinear operator $\N$ defined in \Eqref{newop}, \Eqref{1.6} has the bilinear structure $\N(\V u)=\V B(\V u,\V u)$ with $\V B:\; X_\alpha\times X_\alpha\to L^2_\sigma(\mathscr C)\times\R^3\times\R^3$ defined by 
\begin{equation}\label{eq:b}
\V B(\V u_1,\V u_2):=\V I^{-1}\Big(-2 {\rm P}\,(\V\omega_1\times\V v_2)-{\rm P}\,(\V v_1\cdot\nabla\V v_2)\,,\ -\V\omega_1\times\mathbb I\cdot(\V\omega_2-\V a_2),-\V\omega_1\times\V z_2\Big)^\top
\end{equation}
for every $\V u_1=(\V v_1,\V\omega_1,\V z_1)\,,\ \V u_2=(\V v_2,\V \omega_2,\V z_2)\in X_\alpha$. We recall that the operators $\V I^{-1}$ and the Helmholtz projector ${\rm P}$ are both bounded by \lemmref{1} and \cite[Remark III.1.1]{Galdi11}), respectively. Hence, $\N(\V0)=\N'(\V0)=\V0$. Moreover, since for $\mathscr C$ of class $C^2$ it is (e.g. \cite[\S III.2.3, Lemma III.2.4.3]{Sohr})
$$
\nu^{\frac12}\|\nabla\V v\|_2=\|\V A_0^{\frac12}\V v\|_2\,,\ \ \|\V v\|_\infty\le C\,\|\V A_0^\alpha\|_2\,,\ \ \V v\in {\sf D}(\V A_0)\,,\ \alpha \in (3/4,1)\,,
$$
we easily deduce, for every $\V v$, $\V v_1$ and $\ \V v_2\in Y_\alpha$, and  $\alpha\in (3/4,1)$ that
\be\begin{split}
\norm{{\rm P}(\V v\cdot \nabla \V v)}_2&\le C_1\norm{\V A_0^{1/2} \V v}_2\norm{\V A_0^\alpha\V v}_2\le C_2\norm{\V A_0^\alpha\V v}^2_2,
\\
\norm{{\rm P}(\V v_1\cdot \nabla \V v_1)-{\rm P}(\V v_2\cdot \nabla \V v_2)}_2&\le C_3[\norm{\V A_0^\alpha\V v_1}_2 \norm{\V A_0^{1/2}(\V v_1-\V v_2)}_2+\norm{\V A_0^{1/2} \V v_2}_2\norm{\V A_0^\alpha(\V v_1-\V v_2)}_2]
\\ &\le C_4[\norm{\V A_0^\alpha\V v_1}_2+\norm{\V A_0^\alpha\V v_2}_2]\norm{\V A_0^\alpha(\V v_1-\V v_2)}_2.
\end{split}\eeq{non}
This  shows that the remaining conditions in \Eqref{H4} are satisfied with $p=2$ and $\alpha$ as above. We next pass to the proof of \Eqref{H5}. To this end, let
$$
\V u=\V u^{(0)}+\V u^{(1)}\,,\ \ \V u^{(0)}=(\V v\equiv\V0, \V \omega^{(0)},\V z^{(0)})\in {\sf N}(\L)\,,\ 
\V u^{(1)}=(\V v, \V \omega^{(1)},\V z^{(1)})\in {\sf R}(\L)\,.
$$
Then,  from \Eqref{1.6} we deduce
\be\begin{array}{rl}\medskip
\M(\V u^{(0)}, \V u^{(1)})=\Big(-2{\rm P}[(\V\omega^{(0)}&\!\!\!\!+\V\omega^{(1)})\times\V v]-{\rm P}(\V v\cdot\nabla\V v),\\
 &\!\!\!- (\V\omega^{(0)}+\V\omega^{(1)})\times\T I\cdot(\V\omega^{(0)}+\V\omega^{(1)}-\V a),-(\V\omega^{(0)}+\V\omega^{(1)})\times (\V z^{(0)}+\V z^{(1)})\Big)^\top.
\end{array}
\eeq{mom}
Thus, if $K_A,K_B\neq 1$, from \propref{3.1} we infer
$$
\V\omega^{(0)}\times\T I\cdot\V\omega^{(0)}= \V\omega^{(0)}\times \V z^{(0)}=\V 0\,,
$$  
and \Eqref{H5} follows from the latter, \Eqref{mom} and \Eqref{non}.
\setcounter{equation}{0}
\section{\large Nonlinear Stability Properties}
{ In view of the results obtained in the previous three sections, we are now able to employ the general theory developed in Section 2,  and use \theoref{1} and \theoref{UST} to provide a complete characterization of the stability and asymptotic stability properties of the steady-state motion \eqref{s}. 

\par
We begin to show the following nonlinear stability result. 

\Bt {\rm (Nonlinear stability)}
Let $\alpha\in (3/4,1)$ and assume that the following conditions hold
\be
C>M\equiv\max\{A,B\}\,;\ \ \ {\lambda}^2>\frac{\beta^2}{C-M}\,.
\eeq{sT}
{\color{black} Then,  the steady-state \eqref{s} is stable in $X_\alpha$, namely,
for any $\varepsilon>0$ there is $\delta>0$ such that 
\be
\norm{(\V v_0,\V \omega_0,\V z_0)}_\alpha<\delta\ \ \Longrightarrow \ \ \sup_{t\ge 0}\norm{(\V v(t),\V \omega(t),\V z(t))}_\alpha<\varepsilon\,.
\eeq{St}
Moreover,  there exists $r\in\real$ such that
\be\begin{array}{ll}\medskip\displaystyle{
\lim_{t\to\infty}\|\V A_0^\alpha\V v(t)\|_2=0\,,}\\ \medskip \displaystyle{
\lim_{t\to \infty}\V\omega(t)=r\,\V e_3}\\
\displaystyle{\ 
\lim_{t\to \infty}\V z(t)= \V 0\,,}
 \end{array}
\eeq{ST}
all the above limits occurring at an exponential rate.} 
\ET{6.1}

\begin{proof} By  \theoref{4.3}, we know that, under the stated assumptions \Eqref{sT}, we have $\sigma(\L_1)\subset (0,\infty)$. Therefore, in view of what we have shown in \propref{3.1}--\propref{3.4},   and Section 6,~\footnote{Notice that, in view of \Eqref{sT}, $K_A,K_B\neq 1$.} we may employ \theoref{1} and deduce \Eqref{St} and \Eqref{ST}$_{1,2}$. Furthermore, there exists $z\in\R$ such that 
$$
\lim_{t\to \infty}\V z(t)= z\V e_3\,,
$$
at exponential rate. We now observe that the perturbation field $\V z(t)$ must satisfy \Eqref{zet} (or, equivalently, the data satisfy \Eqref{zet0}). This implies that either $z=2$ or else $z=0$. In the first case, taking into account \Eqref{pert}, we deduce that the terminal state of the coupled system will be of the type $(\V v\equiv\V0,\omega=r \V\gamma,\V\gamma=\e_3)$ which, as it is easily checked, is a steady-state solution to \eqref{eq:motion} corresponding to a constant rigid rotation of $\mathscr S$ around $\V e_3$ with its center of mass $G$ in its {\em lowest} position. However, by the stability property \Eqref{sT}, $G$ must be at all times in a neighborhood of its {\em highest} position. Therefore, we can only have $z=0$, which completes the proof of the theorem.  
\end{proof}
\begin{remark}As an illustration of the results obtained in the previous theorem, consider a ``classical" symmetric top, $\mathscr T$, spinning at sufficiently fast rate around its axis ${\sf a}$ in the vertical direction, ${\sf d}$,  passing through the fixed point $O$ and center of mass $G$ (in its highest position). It is then well known (e.g. \cite[Example 9.7C]{Pars}) that a small disorientation of ${\sf a}$ from ${\sf d}$ will produce a stable  precession  of $\mathscr T$ around ${\sf d}$ with ${\sf a}$ performing small oscillations (nutation). If, however, $\mathscr T$ possess an interior cavity filled up with a viscous liquid, \theoref{6.1} tells us that under the same above circumstances, the axis ${\sf a}$ will eventually reposition itself in the vertical direction ${\sf d}$, at an exponential rate. This fact provides a further example of the stabilizing influence of an interior liquid-filled cavity on the motion of a rigid body \cite{DGMZ,GMpendulum,GMM,Ma,Ma2,MPS,MPS1}. 
\end{remark}

We also have the following instability result, which is an immediate consequence
of  \propref{3.1}, \propref{3.2}, \theoref{4.3}, Section 5, and  \theoref{UST}. 
\begin{theorem}[Nonlinear instability]\label{th:instability}
With the same notation as in \theoref{6.1}, the steady state \eqref{s} is unstable in $X_\alpha$ if either 
 $$C>M\,, \ \ \mbox{and}\ \
\lambda^2<\frac{\beta^2}{C-M}\,. 
$$
or
$$C< M.
$$
\end{theorem}

\ed